# the em—dash em—beds in Congress: a population-level rise in em-dash frequency in U.S. congressional press releases at the dawn of the large-language-model era, 2021–2025

**Przemysław Czuma, MD, MBA** · Polish Association for Artificial Intelligence in Medicine (inteligencja.org.pl) · przemek.czuma@gmail.com · ORCID 0009-0009-8235-2053

## Abstract

**Background.** Large language models (LLMs) can leave small stylistic traces in text written with their help. One commonly discussed trace is the em-dash (—), especially the unspaced English form `word—word`. This form is normal in typeset English prose but unusual in U.S. press writing, where AP style calls for spaces around a dash. This study asks whether the trace can be measured in political communication written by congressional staff.

**Methods.** The study was preregistered in the Open Science Framework (OSF; osf.io/u5ney; DOI 10.17605/OSF.IO/U5NEY) before any confirmatory result was calculated. The corpus was the open *congress-press* dataset of full-text press releases from members of the U.S. House and Senate. To avoid a change in source typography, the analysis used only scraper-sourced records from 2021–2025 (N = 146,239; 480 offices). The primary measure was the density of unspaced prose-form em-dashes (U+2014) per 1,000 characters of cleaned text. Counts were modeled with Poisson/negative-binomial regression, with text length as the exposure and uncertainty adjusted for repeated releases from the same office (Bioguide). The preregistered cut-off was the public release of ChatGPT on 30 November 2022. The plan also included a within-author analysis, a net-new control based on ASCII-hyphen stability, and a mandatory validation gate for removal of the header dateline `CITY, ST —`. That dateline contains a spaced em-dash in every release and would distort a naive measure of em-dash use.

**Results.** Prose-form em-dash density stayed within 0.10–0.12 per 1,000 characters from 2021 through 2024, then rose to 0.217 in 2025, more than twice the four-year baseline. The share of releases containing at least one prose-form em-dash rose from about 13% to 24.8%. Using the calendar-year approximation (post 2023–2025 vs pre 2021–2022), the primary frequency ratio was 1.55 (95% CI 1.28–1.93), just above the prespecified 1.5× threshold. With the exact registered daily cut-off, the ratio was 1.528 and the absolute increase was +0.054 per 1,000; both registered effect-size thresholds were still met. The increase was net-new: ASCII-hyphen density stayed near 2.7 per 1,000. It also appeared within authors: among 262 offices with at least 10 releases in both calendar eras, 75.6% increased their density (median change +0.036 per 1,000; sign test $p \approx 1\times10^{-16}$). In a closed panel of 224 offices present in every year, density rose from 0.103 in 2024 to 0.192 in 2025. Tested artifact explanations were not supported: the extraction pipeline showed no step at the 2024/2025 boundary, and continuing offices, not only newly covered offices, showed the increase. The preregistered residual-dateline-leak threshold was exceeded, but post-hoc sensitivity analyses indicate that this leak does not explain the time trend.

**Conclusions.** In this national corpus of congressional press releases, the frequency of the prose-form em-dash more than doubled in 2025 after four years of stability. The rise was broad, appeared within offices, and was not explained by the measurement artifacts tested here. However, the validation gate was formally breached, so the study's full preregistered decision rule was not met as written. The interpretation is therefore exploratory rather than confirmatory: the pattern is consistent with broad diffusion of LLM-assisted writing, supported by post-hoc sensitivity analyses that limit the dateline-leak concern. The marker is a population-level indicator, not an authorship detector for individual releases. A single change over time also cannot determine whether the 2025 increase was driven only by LLM adoption or partly by staff turnover after the 2024 election.

# 1. Introduction

## 1.1 The em-dash

*(This background is shared with a broader research program. The companion study [1] gives the fuller treatment and primary sources; the present design adapts that study.)*

Punctuation usually attracts little attention. The em-dash is different: it separates a phrase more strongly than a comma but less heavily than parentheses.

Most people outside professional writing use it rarely, partly for a simple mechanical reason: standard keyboards have no em-dash key. U+2014 usually requires a shortcut, autocorrect rule, or menu. Typing one is therefore more deliberate than typing ordinary punctuation.

LLMs have no keyboard cost. They can produce an em-dash as easily as any other token and often use the unspaced English form `word—word`. In recent years, this mark has become an informal sign that readers associate with LLM-generated prose.

That association is still only an anecdote unless it can be measured. This study tests it in a setting not previously measured this way: official U.S. congressional press releases.

## 1.2 From anecdote to measurement

The claim that “LLMs love em-dashes” is common in public discussion but rarely tested with preregistered measurement. Where it has been studied, the direction is similar. Keck reported roughly a doubling of em-dash use in ecology abstracts in OpenAlex between 2021 and 2025 [2]. Liang and colleagues documented machine-modified text at scale in AI conference peer reviews [3]. Kobak and colleagues found over-representation of vocabulary preferred by LLMs in biomedical publications [4]. The companion study found a preregistered increase in em-dash use in Discussion sections of medRxiv preprints [1].

Two conditions make a corpus useful for this question. First, it must preserve the writer's original typography. Second, the human baseline of the marker should be low enough for a change to be visible. Many bibliographic pipelines fail the first condition because they normalize punctuation, sometimes almost completely [5].

Congressional press releases meet both conditions. They are written in English by member offices and are generally published without an outside typesetter, so much of the office's original typography survives. They are also persuasive, narrative texts written for the public. If assisted writing leaves a stylistic trace, this is a plausible place to look. The corpus comes from the open *congress-press* dataset of full-text House and Senate releases [6].

The dataset, however, combines two sources with different typographic behavior: an older API import and a newer scraper pipeline, which meet around 2020. To prevent that source change from looking like a time trend, this study uses only scraper-sourced records from 2021–2025. A pilot integrity check confirmed that the body text preserves the literal U+2014 character, rather than replacing it with ASCII `--`, when compared with the live office page.

## 1.3 Scope of the claim

The research question is deliberately narrow: did the frequency of one punctuation mark change, and when? The confirmatory core (cohort, primary endpoint, main model, prespecified controls and falsification tests, and effect-size thresholds) was frozen and deposited in OSF before any confirmatory number was calculated. Analyses added later are labeled where they appear and collected in §4.6.

The same section also reports an important limitation of the preregistered workflow: the validation gate was formally breached. Therefore, the full preregistered decision rule was not met as written.

This study does **not** build an authorship detector. A human writer may use many em-dashes, and an LLM user may remove every one. The study cannot identify which release involved a model, and it cannot prove that LLMs caused the change. It can only measure whether and when the mark became more common across nearly 150,000 releases, and whether that change coincided in time with wider use of LLM-assisted writing.

## 2. Methods

The full plan is registered in OSF (DOI 10.17605/OSF.IO/U5NEY). The design adapts the earlier preregistered medRxiv study [1] to U.S. congressional press releases: an English-only press-release corpus, a density endpoint, author-clustered models, three preregistered placebo cut-offs, and prespecified effect-size thresholds. Measurement and aggregation used one canonical script, and SHA-256 fingerprints of the frozen dataset were recorded in the registration. All deviations from the plan are listed in §4.6.

### 2.1 Data source and protection against a source seam

The *congress-press* dataset contains one record per release: URL, date, body text, member metadata (Bioguide identifier, party, chamber), and a `date_source` field. It combines a legacy source (≤2020, API import) and a scraper source (2020–present, full-text extraction). Because the two pipelines have different typography, the analysis used only scraper-sourced records throughout 2021–2025.

### 2.2 Cohort and mandatory removal of boilerplate

The cohort included all scraper-sourced releases from members of both chambers in 2021–2025 with non-empty **raw** body text (N = 146,239; 480 offices). The year 2020 was excluded because scraper coverage was too sparse.

Press releases contain a standard header dateline such as `CITY, ST —` and a footer such as ### or social-media links. The dateline itself contains an em-dash, so leaving it in the text would measure the template rather than the writer's prose. The removal rule was frozen in code. A dateline was removed only when its state token matched a closed list of the 50 states, DC, and territories; this conservative rule avoids cutting a normal sentence that starts `Word —`. The footer was removed from the first footer marker onward. Only cleaned body text was measured.

Sixteen releases had non-empty raw text but became empty after boilerplate removal. They remain in N, but contribute zero characters and zero em-dashes to the estimates (§4.6).

### 2.3 Primary endpoint and exposure

The primary endpoint was the density of **unspaced prose-form em-dashes** per 1,000 characters of cleaned text. In plain language, it counts the em-dash placed directly between words, as in `word—word`. Each em-dash was automatically classified as a numeric range (`1934—1971`), a relation between capitalized terms (`Biden—Harris`), or prose (`word—word`). The primary signal was the prose class; ranges and capitalized relations were excluded from M1.

The unspaced form was chosen for four independent reasons.

1. **It was preregistered.** The `word—word` form and the range/relation/prose classification were specified as M1 before em-dashes were counted in this corpus. The earlier medRxiv study [1] measured em-dash presence more generally; the unspaced distinction was introduced here because every congressional release contains a spaced dateline dash.

2. **It has a low human baseline in this genre.** AP style for U.S. press writing calls for spaces on both sides of a dash [7], while Chicago style for typeset prose omits them [8]. The unspaced form is therefore unusual both mechanically (no dedicated key) and stylistically in press writing. By contrast, the spaced form overlaps with normal journalistic typography, including the dateline itself. Empirically, the unspaced form stayed near 0.10 per 1,000 through 2021–2024, while the spaced form had already drifted upward from 0.036 to 0.058 per 1,000.
3. **It is the majority form in the corpus.** Of all U+2014 em-dashes, 67.2% were unspaced, 29.5% spaced, and 3.3% mixed.
4. **The choice does not maximize the observed effect.** Both spaced and unspaced forms rose by a similar relative amount in 2025. The spaced form is reported as a post-hoc sensitivity analysis in §3.5.

Density was chosen instead of simple presence because presence can saturate and is strongly affected by document length.

The preregistered exposure was era, divided at the public release of ChatGPT on 30 November 2022 [9]. For the main manuscript analysis, this was approximated by calendar years: pre = 2021–2022 and post = 2023–2025. December 2022 was therefore assigned to the pre period. The exact daily-cut-off sensitivity analysis is reported in §4.6 (ratio 1.528; absolute increase +0.054 per 1,000). No washout period was used.

## 2.4 Primary analysis and interpretation rule

The primary test modeled the number of prose-form em-dashes by era using Poisson/negative-binomial regression, with an offset for log(cleaned-text length) and cluster-robust standard errors by member (Bioguide) [10]. In plain language, the model does two things: it accounts for the fact that longer releases give more opportunities to use an em-dash, and it adjusts the uncertainty because releases from the same office share a house style and are not independent.

Clustering changes the standard errors, not the point estimate, and it does **not** give every office equal weight. A separate post-hoc analysis tests whether very prolific offices dominate the result (§3.3). Confidence intervals were also checked by bootstrap over offices.

The reporting rule was fixed in advance. The primary estimand was the exposure-weighted post/pre frequency ratio, consistent with the offset model and the modified-Poisson tradition [11]. An increase counted as practically important only if it met **both** thresholds: a ratio of at least 1.5× and an absolute increase of at least 0.05 per 1,000 characters. With a corpus this large, statistical significance alone is not informative enough, which is why the magnitude thresholds were set before analysis.

## 2.5 Supporting, sensitivity, and falsification analyses

The plan specified: a within-author analysis; annual results; a segmented interrupted-time-series (ITS) model to distinguish an immediate step from later acceleration [12]; an en-dash (U+2013) control; and sensitivity to the boilerplate-removal rule.

The preregistered **net-new** control used ASCII-hyphen (U+002D) density. If the increase in em-dashes came mainly from typographic conversion of hyphens into em-dashes, hyphen density should fall. The plan specified that a hyphen decline explaining less than 25% of the em-dash increase would not undermine the net-new interpretation.

The three preregistered **placebo cut-offs** were placed inside the pre-LLM period: June 2021, January 2022, and June 2022. A general fashion-driven upward drift should also produce an apparent increase at one or more false cut-offs.

After the late concentration of the observed increase became known, two additional checks were added: pipeline invariance at the 2024/2025 boundary, and a comparison of continuing offices with offices newly covered in 2025. These are post-registration analyses (§4.6).

### 2.6 Measurement validation: the STOP/GO gate

Because the measurement depends on correct removal of the dateline, the preregistration included a validation gate.

Automatic checks on the full corpus included: unit tests on canonical examples, the annual **residual dateline em-dash rate** (the share of releases in which an em-dash still appears near the start of cleaned text), and a comparison of densities under canonical and looser stripping rules.

A manually adjudicated sample of 300 releases (60 per year) compared raw headers with cleaned text. This manual adjudication was performed after registration and after the census analysis rather than before it, a deviation reported in §4.6. For state-coded datelines (the intended target of the rule), stripping recall was 99.2% (122/123; Wilson 95% CI 95.5–99.9%) [13], with no time trend. No case removed author prose; removed leading material consisted of boilerplate such as "For Immediate Release," contact blocks, and photo captions.

Bare-city datelines without a state code were intentionally left in place to avoid damaging genuine sentences. A positional sensitivity analysis bounded their possible effect. Only 1.4/2.4/1.0/2.4/1.6% of prose-form em-dashes appeared in the first 120 characters in 2021–2025, with no time trend. Recomputing after removing the first 120 or 200 characters left the primary ratio essentially unchanged: 1.554 and 1.550, compared with 1.554 in the main calendar-year analysis.

The gate also required the placebo cut-offs to be negative and the direction of the result to remain unchanged across stripping variants. As reported in §4.6, the **residual-leak thresholds themselves failed**: computed on the full corpus, the automatic leak rate was 7.00 / 4.89 / 3.94 / 4.95 / 4.94% per year in 2021–2025 (weighted mean 4.96%), above the preregistered 2% threshold, and the 2021→2025 change was about 2.1 percentage points, above the 1-point threshold. Therefore, the full preregistered GO ∧ H1–H4 decision rule was not met.

### 2.7 Software and reproducibility

Analyses were run in Python. The code is frozen and version-controlled. Processed per-release measurements and SHA-256 fingerprints will be archived at publication. The *congress-press* corpus is public external data under the MIT license and is not redistributed here. Headline numbers were independently re-derived from the raw frozen snapshot.

## 3. Results

### 3.1 Main result

Across 146,239 releases, prose-form em-dash density was 0.102 per 1,000 characters in 2021–2022 and 0.158 in 2023–2025. The exposure-weighted primary frequency ratio was **1.55 (95% CI 1.28–1.93)**. This is just above the prespecified 1.5× threshold, and the interval excludes 1. The absolute increase was **+0.056 per 1,000 characters**; with the exact daily cut-off it was **+0.054**. Both exceed the prespecified +0.05 threshold.

The change is also easy to see as prevalence. About 24.8% of releases in 2025 contained at least one prose-form em-dash, compared with about 13% in the stable years 2021–2023 (Table 1).

The margin above the practical threshold is thin and is reported as such. The pass/fail result also depends on how the post period is aggregated and on the exact era boundary; §4.6 gives those details. At this sample size the p-value for the main comparison is extremely small but less informative than the prespecified effect size.

### 3.2 2025: sudden acceleration

The annual timing is clear. Density stayed near baseline for four years: 0.0997 in 2021, 0.1035 in 2022, 0.1047 in 2023, and 0.1157 in 2024. It then rose to 0.2167 in 2025. That is about twice the four-year baseline (×2.04). The prevalence rose from about 13% in 2021–2023 to 24.8% in 2025 (Table 1).

The quarterly series makes the timing sharper (Figure 2). Across the sixteen quarters of 2021–2024, density stayed around 0.08–0.13 per 1,000, with a near-zero trend of about +0.001 per quarter. It then rose from 0.130 in 2024Q4 to 0.193 in 2025Q1 and 0.241 in 2025Q2. About 97% of the total increase occurs in 2025.

The preregistration promised a segmented ITS analysis at the November 2022 cut-off. The document-level version was fitted during revision, after the descriptive results were known, and used a different unit from the quarterly model described in the registration. Both deviations are disclosed in §4.6. This document-level model used Poisson regression with a log-length offset and clustering by office. It included **146,132 releases**, not the full 146,239: 107 were excluded, consisting of 16 releases empty after boilerplate removal and 91 without an assigned author; no included/excluded difference was caused by a missing daily date.

The document-level ITS showed: a flat pre-cut-off trend (×1.01/year, 95% CI 0.81–1.26); no upward step at the cut-off (step ratio 0.76, 95% CI 0.60–0.96, if anything a temporary dip); a post-cut-off slope of ×1.45/year (95% CI 1.32–1.59); and a slope-change ratio of ×1.43 (95% CI 1.11–1.85). In other words, the model supports acceleration after the cut-off, not an immediate upward step at it.

A second ITS model used the quarterly unit described in the preregistration (20 quarterly points, total character exposure as the offset, quasi-Poisson). It agreed with the document-level version in every term: pre trend ×0.98/year (95% CI 0.72–1.34), step 0.78 (0.53–1.15), post slope ×1.45/year (1.32–1.60), and slope change ×1.48 (1.07–2.05).

The preregistered line dividing the main eras remains 30 November 2022. The time series adds a more precise description: there was no immediate rise at that date, and almost all of the observed increase arrived in 2025. A possible delayed-diffusion interpretation and the competing explanation of the new 119th Congress are discussed in §4.3.

### 3.3 The increase is net-new and within-author

Two controls support the reality of the change.

First, ASCII-hyphen density stayed stable: 2.75 / 2.58 / 2.79 / 2.76 / 2.68 per 1,000 characters from 2021 through 2025. Post-period hyphen density was not lower than pre-period density. Under the preregistered criterion, conversion from hyphens therefore explains 0% of the em-dash increase, well below the 25% limit. The result is more consistent with additional em-dash use than with mass conversion of hyphens into em-dashes.

Second, the increase appeared within offices. Among **262 offices** with at least 10 releases in both calendar eras, **75.6%** increased their em-dash density; the median change was **+0.036 per 1,000**, and the office-level sign test gave $p \approx 1\times10^{-16}$.

The result was not driven by a few very prolific offices. In a post-hoc analysis, removing the 24 most prolific offices (top 5% by release count, representing 15.7% of the corpus) left the exact-cut-off ratio intact and numerically higher: **1.66 (95% CI 1.31–2.09)**, compared with **1.528** in the full sample under the same daily-cut-off specification.

A closed panel of **224 offices** present in every year showed the same pattern: 0.100 / 0.091 / 0.107 / 0.103 / 0.192 from 2021 through 2025. The same offices changed their style in 2025; the pattern is not only the result of new offices entering the dataset.

### 3.4 Alternative artifact explanations were not supported

A large change concentrated in one year could be an artifact, so two simple explanations were tested directly.

**Pipeline invariance.** If the 2025 increase came from a change in text extraction, the extraction pipeline should also show a break at the 2024/2025 boundary. It did not. The dominant scraper value of `date_source` stayed at about 99% on both sides of the boundary; the share of releases with a detectable canonical dateline stayed around 48–51% in nearby quarters (41% state-coded datelines in the year-stratified validation sample); and median length and residual leak showed no boundary step. These checks do not support an extraction-change explanation.

**New versus continuing offices.** Corpus coverage grows over time, and 2025 has the most releases. But offices already present in every year from 2021–2024 (n = 225, essentially the stable panel above) increased from 0.103 in 2024 to 0.192 in 2025. Offices first covered in 2025 had a higher density, 0.267, but they only added to a rise already present in continuing offices.

The three preregistered placebo cut-offs inside the pre-LLM window were also flat when tested with the main document-level Poisson specification:

| False cut-off | RR | 95% CI |
|---|---:|---:|
| June 2021 | 0.989 | 0.812–1.205 |
| January 2022 | 1.058 | 0.843–1.329 |
| June 2022 | 0.941 | 0.720–1.230 |

None shows an increase. The tested extraction and composition artifacts were therefore not supported.

### 3.5 Robustness

The direction of the result was unchanged under the canonical versus looser stripping rule; annual densities differed only by a fraction of a percent. Including the `relation` class did not change the direction. The en-dash control remained flat. Annual and within-author results were independently replicated from the frozen snapshot.

The result also did not depend on spacing. The spaced form (`word — word`) has a noisier and already rising baseline, but it still rose from about 0.044 per 1,000 in 2021–2023 to 0.096 in 2025, roughly a doubling. This spacing analysis was added after registration (§4.6). Thus, the 2025 discontinuity appears in both spaced and unspaced forms.

# 4. Discussion

## 4.1 What this study showed

In this national corpus, prose-form em-dash density in congressional press releases rose in 2025 to more than twice its earlier stable level. The increase was net-new, appeared within authors and in a closed panel of the same offices, and survived checks designed to detect changes in extraction or sample composition.

One possible interpretation is that these independent patterns reflect wider use of LLM-assisted writing as models matured. That interpretation fits the timing, but the study does not identify a cause.

## 4.2 What the result means, and what it does not mean

This is a population-level change, not a judgment about an individual release. The marker describes a distribution across nearly 150,000 documents. It cannot tell who wrote a specific text. Some people naturally use em-dashes, and some LLM users remove them.

The narrow result is stronger: the style of congressional press releases changed in 2025, broadly and in a way not explained by the tested measurement artifacts. The change coincides in time with wider use of LLM-assisted writing. That is a statement about a corpus-level writing pattern, not proof of individual authorship or causation.

## 4.3 Why 2025? The delayed-diffusion hypothesis

One possible interpretation is that the marker increased not when LLMs first became publicly available, but when the models became good enough for routine institutional work. The 2022 generation could write fluent text, but it was often unreliable, verbose, and easy to recognize. Models in 2024–2025 became more capable, accurate, and natural. Under this interpretation, public availability in 2022 was a necessary early condition, while later improvements made assisted drafting, editing, shortening, and translation practical enough for regular use. The study cannot establish this mechanism; it can only say that the observed timing is compatible with it.

There is an important competing explanation. January 2025 was also the start of the 119th Congress, a new administration, unified single-party control, and a change in the Senate majority. Staff turnover or a new political communication climate could therefore change writing style without LLM diffusion.

An exploratory stratified analysis added after registration tests part of this concern. The 2025 increase was similar across parties and chambers. Within continuing offices, the 2025/pre fold change was about 1.9× for both Democrats and Republicans. The within-office ratio of these fold changes was **0.96 (95% CI 0.67–1.38)**, including 1. The population Senate/House ratio of fold changes was **0.63 (95% CI 0.36–1.09)**; the House rose more, but the interval also included 1. Both parties rose already in 2025Q1. The small remaining party difference favored Democrats: Hodges–Lehmann shift R–D **−0.041 per 1,000 (95% CI −0.071 to −0.010)**; Mann–Whitney p = **0.009**. This is the opposite of a simple story in which the incoming majority alone drives the change.

A separate exploratory lexical check points in the same general direction, but not with identical timing. The document-level post/pre odds ratio was **1.57 (95% CI 1.39–1.76)** for a composite of the three most frequent LLM-preferred words from Kobak et al. [4], and **2.07 (95% CI 1.76–2.44)** for *underscore* alone. That lexical signal began to build in 2023→2024, earlier than the em-dash jump, so the convergence is directional rather than exact.

These exploratory results make a simple party-composition explanation less convincing, but they do not observe tool use itself. Whether new staff brought the habit or continuing staff adopted new tools remains unresolved.

**The corpus measures a marker, not LLM use.** Delayed diffusion as models matured is therefore an interpretation that fits the trajectory, not a demonstrated mechanism (§4.6).

### 4.4 Why this matters, especially in politics

A congressional press release is part of an elected representative's official public voice. Offices use releases to state positions, respond to events, criticize opponents, and communicate with constituents. If machine-assisted writing becomes common in this channel, questions of authenticity, responsibility, transparency, and possible homogenization of political language become relevant.

The study does not show that any specific release is "fake," and it does not show that a machine replaced a member of Congress or staff writer. It shows a narrower fact: the style of official political communication changed measurably and quickly. That is enough to raise a policy question (whether public officials should disclose AI assistance in official communication), but this study does not answer that policy question.

Congress is also a particularly useful place to measure the signal because original typography is well preserved. The study does not establish that the same phenomenon occurs outside U.S. congressional press releases; that broader question requires separate data.

### 4.5 Limitations

First, and most important, the em-dash is **not a detector**. It has no reliable meaning for an individual release and does not support a causal claim.

Second, the main increase is concentrated in 2025, which also marks the start of a new Congress, a new administration, and unified single-party control (§4.3). The stratified analysis shows symmetry across parties and within offices, which weighs against a simple political-composition mechanism. Even so, this corpus cannot fully separate LLM diffusion from changes after the 2024 election.

Third, the text comes from a third-party extraction pipeline. Validation confirmed fidelity of the em-dash to live pages and pipeline stability around the 2024/2025 boundary, but the full version history of the extractor is unavailable.

Fourth, some script-rendered pages were unavailable, more often for Republican offices early in the period. Missing text was low (0.4–0.8%) and is reported. The within-author analysis reduces concern about this imbalance, but it does not remove uncertainty about absolute levels.

Fifth, this is one country and one genre. Whether the same trajectory appears elsewhere is an open empirical question.

### 4.6 Deviations from the preregistration

Five groups of deviations must be disclosed. None removes the observed empirical increase, but they are not cosmetic. The first two directly affect the thin margin around the primary practical-effect threshold. The third means that the full preregistered rule **GO ∧ H1–H4 was not met as written**. For that reason, the interpretation "consistent with LLM adoption" is exploratory and supported by post-hoc sensitivity analyses, not confirmed by the registered gate.

**1. Aggregation of the primary ratio.** The registration did not fully specify how to combine yearly densities into the post era: a pooled density (total em-dashes / total characters) or a simple average of annual densities. The pooled/model estimand is consistent with the density definition and the log-length offset model and gives **1.55×**.

A naive average of annual densities gives **1.43×**, and its absolute increase is only **+0.044 per 1,000**. Under the yearly-average approach, the practical-significance criterion fails both thresholds.

The difference is caused by exposure weighting: 2025, the year with the largest increase, also has by far the largest coverage (**48,489 releases**, compared with **30,222 in 2023**). The manuscript uses the pooled/model estimate as primary because it matches the registered offset-model specification and the registration's statement that the five annual points are not the primary test. However, the registration defines M1 “per 1,000 characters of cleaned body, per year” and gives an explicit Σ em-dashes / Σ characters form only for the per-author estimator in H3. It does **not** explicitly settle era-level aggregation. That ambiguity is why this choice is disclosed as a deviation. The honest summary is: H1 passes under the pooled estimate (**1.554** by calendar years; **1.528** at the exact daily cut-off) and fails under the simple yearly average.

**2. Era boundary.** The registration defined pre = 2021-01..2022-11 and post = 2022-12..2025-12, using 30 November 2022 as the daily cut-off. The main manuscript approximates the periods by full calendar years, placing December 2022 in the pre era. December 2022 contained **1,907 releases** with density **0.088**, below the 2022 average, so this assignment is mildly conservative. With the exact registered cut-off, the main ratio is **1.528** and the absolute increase is **+0.054 per 1,000**; both thresholds remain met, with an even thinner margin.

The calendar approximation also affects all era-based analyses, including H3 and Figure 3. With the exact cut-off, H3 gives **260 offices**, **75.4%** increasing, median Δ **+0.034 per 1,000**, sign-test p ≈ **$9.5\times10^{-17}$**, compared with **262 / 75.6% / +0.036** under calendar-year periods. The conclusion is unchanged.

**3. Dateline validation and the residual-leak gate.** The preregistration planned a manually coded validation sample as a gate **before** the full census. The 300-release sample (60/year) was adjudicated after registration and after the census, so it is reported as post-hoc rather than as a gate executed in the planned order. The automatic leak check available when the census proceeded had also used the looser, crude stripping rule, which understates leakage; the values reported here come from the canonical recomputation.

Manual adjudication was strong: recall for state-coded dateline stripping was **99.2% (122/123)**, stable across years, with no author prose removed (§2.6). But the automatic residual-leak rate was **7.00 / 4.89 / 3.94 / 4.95 / 4.94% per year in 2021–2025 (range 3.9–7.0%; weighted mean 4.96%), computed on the full corpus**, above the preregistered **2%** threshold. Its 2021→2025 change was about **2.1 percentage points**, above the registered **1-point** threshold. This is a formal gate failure. (The year-stratified validation months give slightly different values, for example 6.4% in 2021, because the base differs.)

Three results limit the practical effect of this failure. First, most leakage comes from bare-city datelines such as `WASHINGTON —`, which the conservative rule intentionally leaves. Their em-dash is spaced and therefore outside the unspaced primary M1. The positional sensitivity analysis directly bounds any remaining unspaced contribution and leaves the main ratio unchanged. Second, the leak moves against the em-dash trend: it is highest in the baseline year and lower in the years of the rise (**7.00%→4.94%**); it fell to **3.94% by 2023 and rose only modestly to 4.94%** in 2023→2025; a leak-driven artifact would require the opposite pattern. The important mitigation is the spacing classification and positional sensitivity, not the direction of the leak trend. Third, annual prose-form densities are almost identical under the canonical and crude stripping variants. The stripping rule therefore does not appear to drive the time trend, but the gate breach remains a disclosed deviation.

**4. Exploratory analyses added after registration and an ITS model-form deviation.** Because the increase was unexpectedly concentrated late, two discriminating tests were added after results were known: pipeline invariance and continuing-versus-new offices. Other post-registration additions were the January-2025 alternative cut-off (**about 2.0×** for 2025 vs pooled 2021–2024), the spaced-form and spacing-share analyses, party × era and chamber × era stratification, the lexical analysis based on the three most frequent Kobak-list words [4], and the positional sensitivity analysis in §2.6.

The promised segmented ITS also differed from the registration in two ways: it was fitted at revision after descriptive results were known, and the first fitted version used release-level data rather than the quarterly regression described in the plan. Both are disclosed in §3.2. A quarterly model using the registered unit was then fitted as a control and agreed with every term of the document-level model. These analyses are labeled exploratory or post-hoc, not part of the frozen confirmatory plan. Where no confidence interval is given (the descriptive 2.0× January-2025 contrast, pipeline-composition values, and spacing shares), the number should be read descriptively.

**5. Registered items not executed.** For completeness, the following registered items were not completed:

- The `langid` filter, p(English) ≥ 0.80, was not applied. The corpus is institutionally English, and no non-English release appeared in any validation sample.
- The registered content-hash deduplication sensitivity analysis was not run. Deduplication used URL; there were zero duplicate URLs in the frozen snapshot.
- Differential missingness was analyzed only by party (missing body text 0.4–0.8%). The planned year × chamber × party breakdown and author-level missingness sensitivity were not produced.
- The M2 model with document length as a covariate and the all-em-dash-form lower/upper bounds were not reported. Median document length increased from **2,099 to 2,377 characters**, so M2 is treated as descriptive.
- The planned year-specific precision/recall estimates for stripping were not available. The adjudicated sample supports only the pooled estimate 122/123 because there were too few state-coded cases for stable annual estimates.
- Of the remaining annual gate diagnostics, the amount of text removed was calculated and was **0.09/0.04/0.06/0.03/0.02% of characters** by year, with no trend that could create the signal. The planned annual full-corpus dateline-regex match rate was not produced; it is reported instead for quarters around the 2024/2025 boundary (§3.4) and for sampled validation months.
- The manual sample was stratified by year but not additionally by chamber, as the plan specified.
- The **16 releases** that became empty after boilerplate removal remain in N. They contribute no characters and no em-dashes to any estimate.

These omissions do not change the primary pooled estimate, but they are listed so that the registered and executed analyses can be compared item by item.

## 4.7 Conclusions

In U.S. congressional press releases, prose-form em-dash frequency more than doubled in 2025, more than two years after the public release of ChatGPT.

The empirical increase was shown under the frozen core analysis: H1–H4 passed, and the falsification checks did not find a tested artifact. But the validation gate was formally breached, so the full preregistered decision rule was not met. The interpretation is therefore exploratory. The effect-size thresholds were set in advance, and deviations from the plan are listed individually in §4.6.

An em-dash cannot determine whether a specific release was written with an LLM. At the population level, it does show a large, broad, and clearly timed change in how congressional offices write. The timing is itself important: the change appeared substantially later than the simplest “ChatGPT released, writing changed immediately” story would predict.

## Tables

**Table 1. Annual prose-form em-dash density and prevalence.**

| Year | Density /1,000 characters | Releases with ≥1 prose em-dash |
|---|---:|---:|
| 2021 | 0.0997 | 12.9% |
| 2022 | 0.1035 | 12.9% |
| 2023 | 0.1047 | 13.4% |
| 2024 | 0.1157 | 14.7% |
| 2025 | 0.2167 | 24.8% |

**Table 2. Main contrasts and punctuation controls.**

| Measure | Result |
|---|---:|
| Primary frequency ratio, post (2023–2025) / pre (2021–2022) | 1.55 (95% CI 1.28–1.93) |
| 2025 vs pre | 2.13× |
| ASCII hyphen /1,000, 2021–2025 | 2.75 / 2.58 / 2.79 / 2.76 / 2.68 |
| En-dash /1,000, 2021–2025 | 0.28 / 0.35 / 0.31 / 0.29 / 0.28 |

**Table 3. Within-author and stable-panel results.**

| Measure | Result |
|---|---:|
| Offices with ≥10 releases in both calendar eras | 262 |
| Offices with increased density | 75.6% |
| Median within-office Δ | +0.036 /1,000 |
| Sign test | $p \approx 1\times10^{-16}$ |
| Stable panel, offices present every year | 224 |
| Stable-panel density, 2021–2025 | 0.100 / 0.091 / 0.107 / 0.103 / 0.192 |

**Table 4. Quarterly em-dash trajectory.** Density per 1,000 characters; N = number of releases. *Unspaced* densities come from the canonical snapshot. *Spaced* counts come from the per-dash context file generated with the looser (crude) stripping variant; the variants agree on annual density to a fraction of a percent (§3.5).

| Quarter | N | Unspaced | Spaced | Quarter | N | Unspaced | Spaced |
|---|---|---|---|---|---|---|---|
| 2021Q1 | 4,214 | 0.104 | 0.026 | 2023Q3 | 6,880 | 0.095 | 0.057 |
| 2021Q2 | 4,185 | 0.104 | 0.033 | 2023Q4 | 7,432 | 0.101 | 0.049 |
| 2021Q3 | 4,189 | 0.088 | 0.040 | 2024Q1 | 8,769 | 0.114 | 0.059 |
| 2021Q4 | 3,693 | 0.104 | 0.047 | 2024Q2 | 9,474 | 0.105 | 0.065 |
| 2022Q1 | 5,030 | 0.105 | 0.049 | 2024Q3 | 7,897 | 0.121 | 0.055 |
| 2022Q2 | 5,107 | 0.117 | 0.039 | 2024Q4 | 5,428 | 0.130 | 0.048 |
| 2022Q3 | 5,218 | 0.108 | 0.042 | 2025Q1 | 13,775 | 0.193 | 0.074 |
| 2022Q4 | 4,324 | 0.081 | 0.041 | 2025Q2 | 12,784 | 0.241 | 0.104 |
| 2023Q1 | 7,898 | 0.111 | 0.044 | 2025Q3 | 11,169 | 0.215 | 0.099 |
| 2023Q2 | 8,012 | 0.111 | 0.045 | 2025Q4 | 10,761 | 0.218 | 0.107 |

## Figures

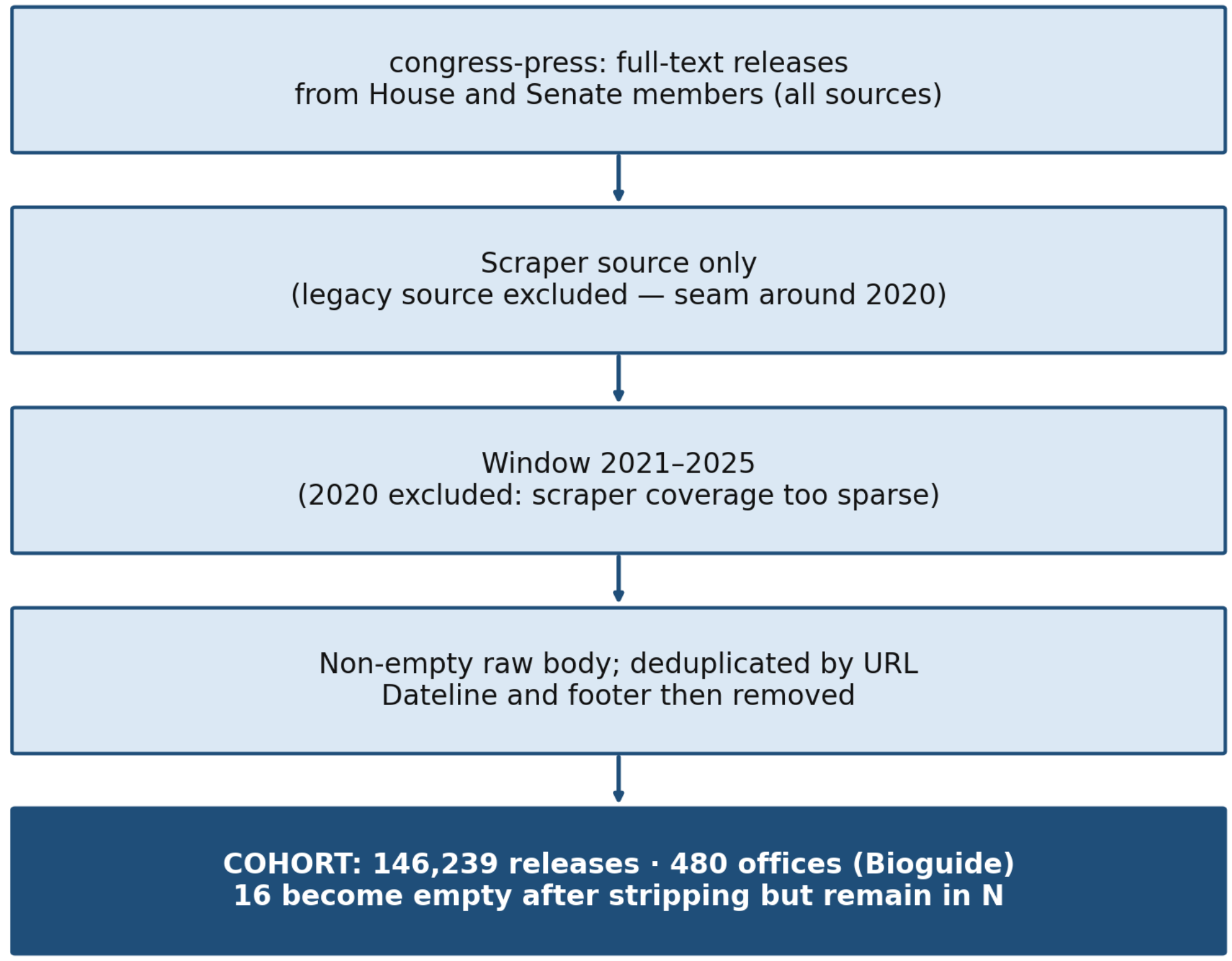

**Figure 1. Study flow.** Full *congress-press* release set → scraper source only (legacy source seam around 2020 excluded) → 2021–2025 window → non-empty raw body → cohort of 146,239 releases from 480 offices. Sixteen releases become empty after boilerplate removal but remain in N and contribute zero characters and zero em-dashes (§4.6).

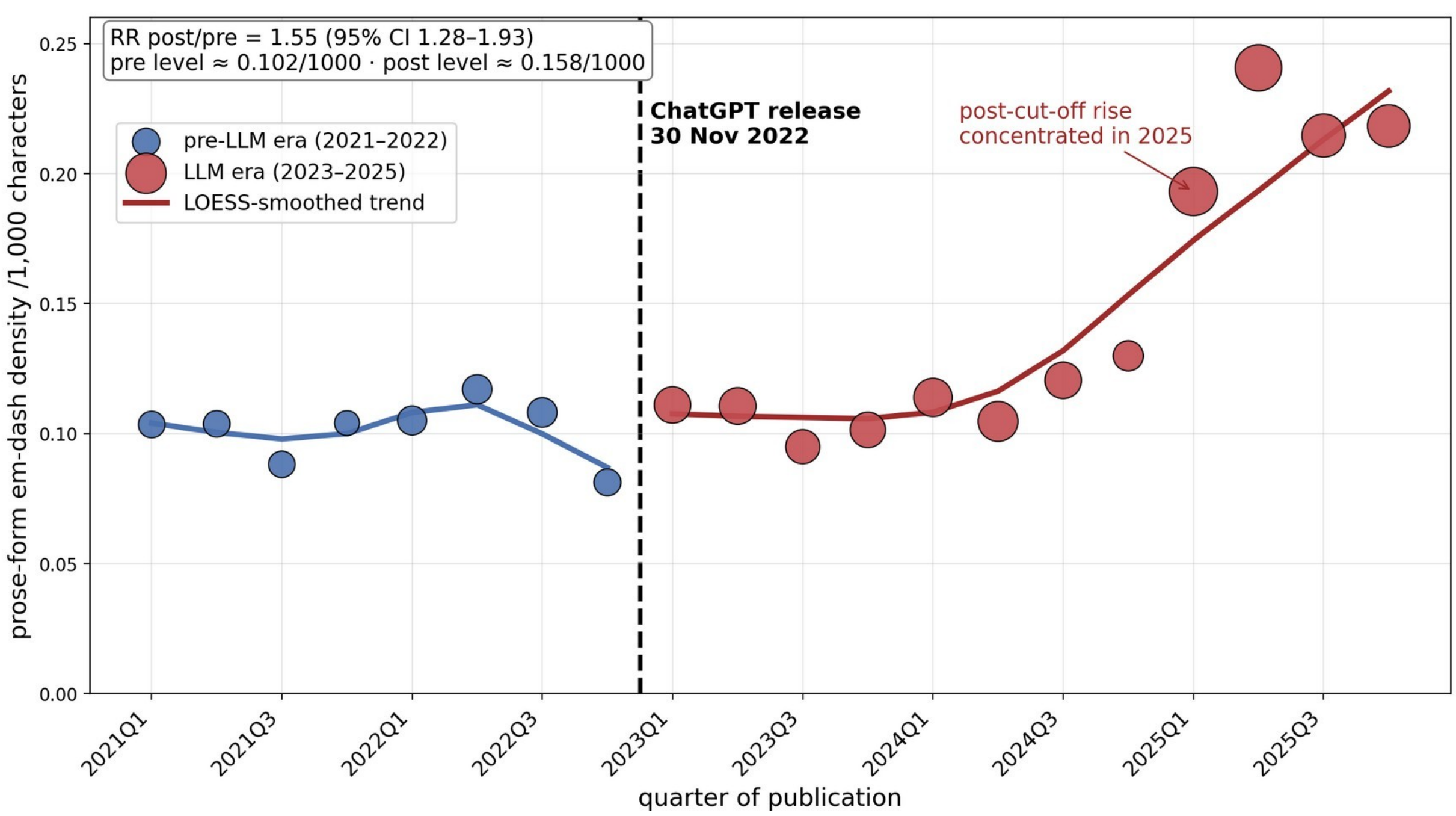

**Figure 2. Quarterly trajectory of unspaced prose-form em-dash density per 1,000 characters.** Twenty quarters, 2021Q1–2025Q4; point size is proportional to the number of releases. The black dashed line marks the preregistered ChatGPT cut-off (30 November 2022). Points are colored as pre-LLM (2021–2022) and post-cut-off (2023–2025), with a LOESS-smoothed trend. The curve stays flat through 2024 and turns upward in 2025 (§4.3).

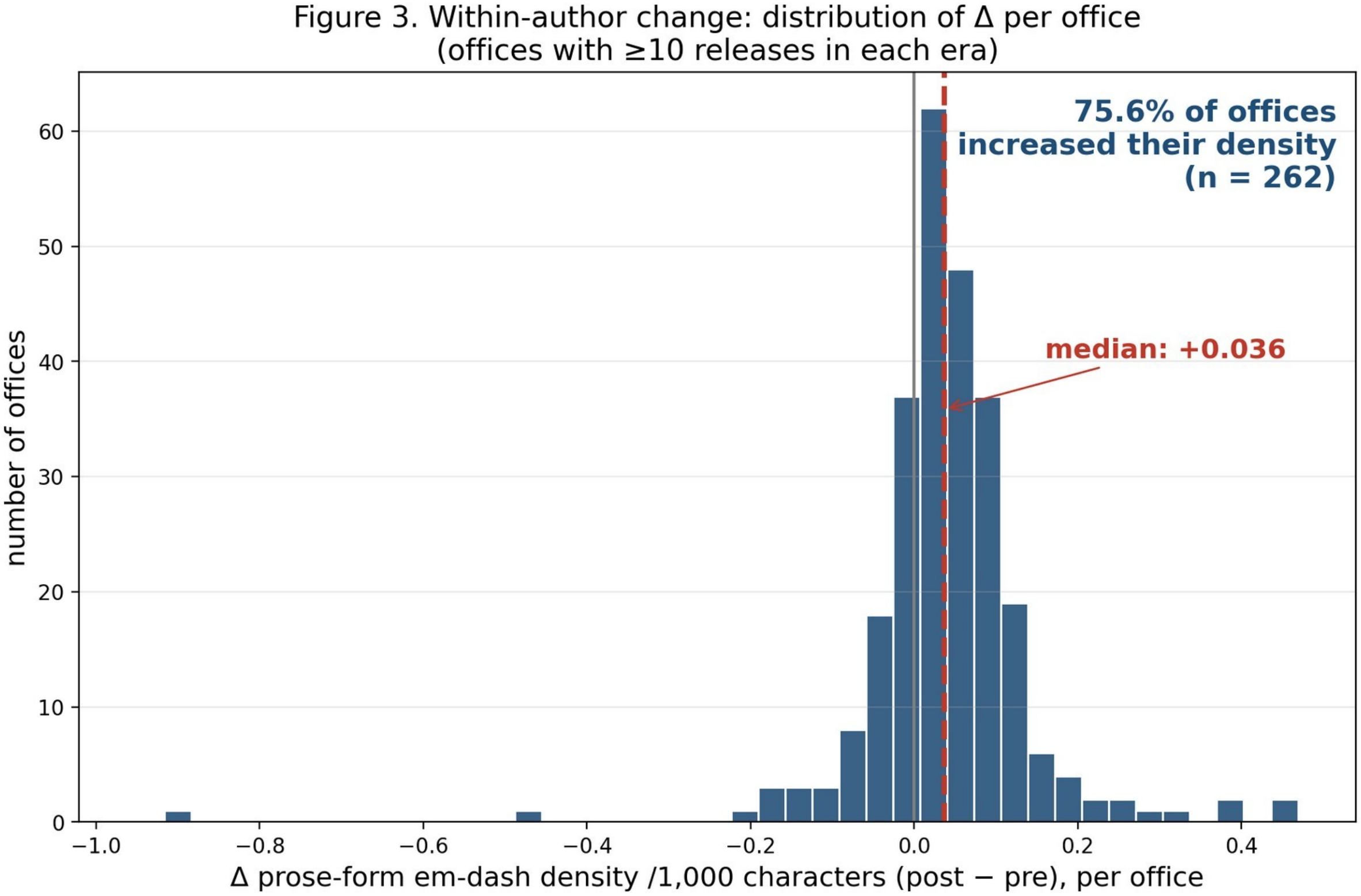

**Figure 3. Within-author change.** Distribution of the office-level change in prose-form em-dash density (post 2023–2025 minus pre 2021–2022) among 262 offices with at least 10 releases in each calendar era. Overall, 75.6% of offices increased their density; median Δ = +0.036 per 1,000.

## Declarations

**Relationship to previous work.** This study is part of a program measuring the same marker (the em-dash as a population-level trace of LLM-assisted writing) in different corpora. The design was adapted from the author's earlier preregistered medRxiv study [1]. The two studies use different data and report different results (medRxiv vs U.S. Congress), so this is not duplicate publication. The shared methodology is explicitly cited. This manuscript uses original wording and does not reproduce the companion text verbatim.

**Data and code.** *congress-press* is a public MIT-licensed dataset. Code and the frozen snapshot (SHA-256 recorded in the registration) will be deposited on Zenodo/GitHub at publication.

**Competing interests.** None.

**Funding.** None.

**AI-use disclosure.** An AI assistant was used for helper code, debugging, language editing, and analytic support under supervision. It did not formulate the research question, choose the corpus, or make autonomous scientific interpretations.